\documentclass[sigconf]{acmart}

\usepackage{booktabs} 
\usepackage{IEEEtrantools}
\graphicspath{{./Bilder/}}

\usepackage{amsmath}
\usepackage{amssymb}
\usepackage{nicefrac}

\usepackage{tikz}
\usetikzlibrary{arrows}    
\usepackage{circuitikz}
\usepackage{subfig}
\usetikzlibrary{circuits.ee.IEC}
\usetikzlibrary{shapes.geometric,calc}

\usepackage{pgfplots} 							
\pgfplotsset{compat=newest}						
\pgfplotsset{plot coordinates/math parser=false} 		
\usepgfplotslibrary{colormaps}

\usepackage{bm}

\usepackage{comment}
\usepackage{siunitx}
\usepackage{epstopdf}

\usepackage{changebar}

\usepackage{siunitx}

\setcopyright{rightsretained}

\acmDOI{10.475/123_4}

\acmISBN{123-4567-24-567/08/06}

\acmConference[NANOCOM'18]{ACM/IEEE NanoCom 2018}{Sept. 2018}{Reykjavik, Iceland}
\acmYear{2018}
\copyrightyear{2018}

\acmPrice{15.00}

\begin{document}
\title[Biological Optical-to-Chemical Signal Conversion Interface]{Biological Optical-to-Chemical Signal Conversion Interface: \\ A Small-scale Modulator for Molecular Communications}

\author{%
Laura~Grebenstein\IEEEauthorrefmark{1}$^\star$,
Jens~Kirchner\IEEEauthorrefmark{2}$^\star$,
Renata~Stavracakis~Peixoto\IEEEauthorrefmark{1},
Wiebke~Zimmermann\IEEEauthorrefmark{1},
Wayan~Wicke\IEEEauthorrefmark{3},
Arman~Ahmadzadeh\IEEEauthorrefmark{3},
Vahid~Jamali\IEEEauthorrefmark{3},
Georg~Fischer\IEEEauthorrefmark{2},
Robert~Weigel\IEEEauthorrefmark{2},
Andreas~Burkovski\IEEEauthorrefmark{1},
and Robert~Schober\IEEEauthorrefmark{3}
}
\affiliation{%
\institution{\IEEEauthorrefmark{1}Institute for Microbiology, 
\IEEEauthorrefmark{2}Institute for Electronics Engineering,
\IEEEauthorrefmark{3}Institute for Digital Communications}
\institution{Friedrich-Alexander-University of Erlangen-Nuremberg}}

\thanks{$^\star$ Co-first authors}

\renewcommand{\shortauthors}{Grebenstein and Kirchner \textit{et al.}}

\begin{abstract}
Although many exciting applications of molecular communication (MC) systems are envisioned to be at microscale, the available MC testbeds  reported in the literature so far are mostly at macroscale. This may partially be due to the fact that controlling an MC system at microscale is quite challenging. To link the macroworld to the microworld, we propose a biological signal conversion interface that can also be  seen as a microscale modulator. This interface translates an optical signal, which can be easily controlled using a light-emitting diode (LED), into a chemical signal by changing the pH of the environment. The modulator is realized using \textit{Escherichia coli} bacteria that express the light-driven proton pump gloeorhodopsin from \textit{Gloeobacter violaceus}. Upon inducing external light stimuli, these bacteria can locally change their surrounding pH level by exporting protons into the environment. Based on measurement data from a testbed, we develop an analytical model for the induced chemical signal  as a function of the applied optical signal. Finally, using a pH sensor as detector, we show for an example scenario that the proposed setup is able to successfully convert an optical signal representing a sequence of binary symbols  into a chemical signal with a bit rate of $1$~bit/min.
\end{abstract}


 \maketitle
\section{Introduction}

Molecular communication (MC) systems encode information into the characteristics of signaling molecules. This is very different from  conventional electromagnetic- (EM-) based  communication systems that embed data into the properties of EM waves \cite{Nakano_Molecular_2013,Farsad_comprehensive_2016}. MC systems are suitable for communication at small scale and in fluids where  EM-based communication is inefficient or even infeasible. Functioning MC systems are envisioned to enable revolutionary applications, e.g., sensing of a target substance  in 
biotechnology,   targeted drug delivery in medicine, and  monitoring of oil pipelines or chemical reactors in industrial applications. 

An important step towards realizing the aforementioned applications is to build testbeds that allow the verification of the theoretical channel models and the transmission strategies proposed in the MC literature. To this end, MC testbeds based on spraying alcohol  into open space and  using acids and bases within closed vessels have been proposed in \cite{Farsad_Tabletop_2013} and \cite{Farsad_Novel_2017}, respectively. These testbeds have been extended to multiple-input multiple-output (MIMO) systems, and improved channel models have been proposed to account for discrepancies between theory and experimental results \cite{Koo_Molecular_2016,Farsad_Channel_2014}. Recently,  an in-vessel MC testbed was proposed in \cite{TestBed_Harold} that uses specifically
designed magnetic nanoparticles as information carriers, which are biocompatible, clinically
safe and do not interfere with chemical processes like alcohol \cite{Farsad_Tabletop_2013} or acids and bases \cite{Farsad_Novel_2017}  
may do. Nevertheless, the aforementioned MC testbeds are all at macroscale, i.e., with dimensions on the order of several tens of centimeters, whereas many prospective applications of MC systems are envisioned to be at microscale. Biologically inspired experimental studies have been conducted in \cite{Krishnaswamy_Time_2013,Nakano_Microplatform_2008,Felicetti_Modeling_2014,Akyildiz_testbed_2015,Nakano_Interface_2014}. In particular, in  \cite{Krishnaswamy_Time_2013}, bacterial populations were used as transceivers connected through a microfluidic pathway.  In \cite{Felicetti_Modeling_2014}, soluble CD40L molecules were released from platelets (as transmitter) into a fluid medium that upon contact triggered the activation of endothelial cells (as receiver). Moreover, in \cite{Nakano_Microplatform_2008}, a microplatform was designed  to demonstrate the propagation of molecular signals through a line of patterned HeLa cells (human cervical cancer cells) expressing gap junction channels. In \cite{Nakano_Interface_2014}, artificially synthesized materials were embedded into the cytosol of living cells and, in response to stimuli induced in the cell, emitted fluorescence that could be externally detected by fluorescence microscopy. Similarly, in \cite{Akyildiz_testbed_2015}, the response of genetically engineered \textit{Escherichia coli} (\textit{E. coli}) bacteria to the surrounding molecules was used as basis for the design of a biological receiver. 

One particular challenge for designing microscale MC testbeds is the fact that controlling an MC system at microscale is difficult. To address this issue, in this paper, we propose a biological signal conversion interface  which converts an optical signal, which can be easily controlled using a light-emitting diode (LED), into a chemical signal by changing the pH of the environment. This setup can be  seen as a microscale modulator that can be embedded in future MC systems\footnote{Throughout the paper, we use the terms ``optical-to-chemical signal converter'' and ``modulator'' interchangeably.}. The modulator is realized using \textit{E. coli} bacteria that express the light-driven proton pump gloeorhodopsin (GR), a bacterial type~I rhodopsin. Upon inducing external light stimuli, these bacteria can change their surrounding pH level by exporting protons into the environment. The authors of \cite{Choi_Cyanobacterial_2014} examined the proton flux due to illumination of \textit{E. coli} bacteria expressing GR (but not for applications in an MC system). In particular, one proton can be transferred to the periplasmic space in less than $1$~ms from an almost inexhaustible pool inside the cell arising from the cell's energy metabolism \cite{Lanyi_Proton_2006}. As a result, in a bacterial suspension, the change of proton concentration in the surrounding medium can be detected within few seconds as a change of pH. Therefore, we expect a relatively fast signal conversion with this setup in comparison with the setup in \cite{Krishnaswamy_Time_2013} where a chemical signal was generated based on gene expression. Using experimentally derived data from our testbed, we develop an analytical model for the induced chemical signal  as a function of the applied optical signal. Finally, using a pH sensor as detector, we show for an example scenario that the proposed setup is able to successfully convert an optical signal representing a sequence of binary symbols into a chemical signal with a bit rate of $1$~bit/min using on-off keying (OOK) modulation and differential detection. The proposed setup can serve as the basis for the development of testbeds using other light-driven pumps that generate other chemical signals, e.g., Na$^+$ and K$^+$ ions~\cite{LightPump_2013,LightPump_2017}. 

We note that the systems in \cite{Nakano_Microplatform_2008,Felicetti_Modeling_2014,Nakano_Interface_2014,Akyildiz_testbed_2015} were demonstrated for a single shot transmission. Furthermore, the setup with continuous transmission in \cite{Krishnaswamy_Time_2013} achieves low data rates on the order of one bit/h. In contrast, the testbed in this paper achieves significantly higher data rates on the order of one bit/min. 


\section{System Setup and Preliminaries}
In this section, first, an overview of the experimental system is provided. Subsequently, the photocycle of bacteriorhodopsin, the main  biological mechanism that is exploited for the proposed microscale modulator, is discussed in  detail.

\begin{figure}[t]
    \includegraphics[width=0.82\columnwidth]{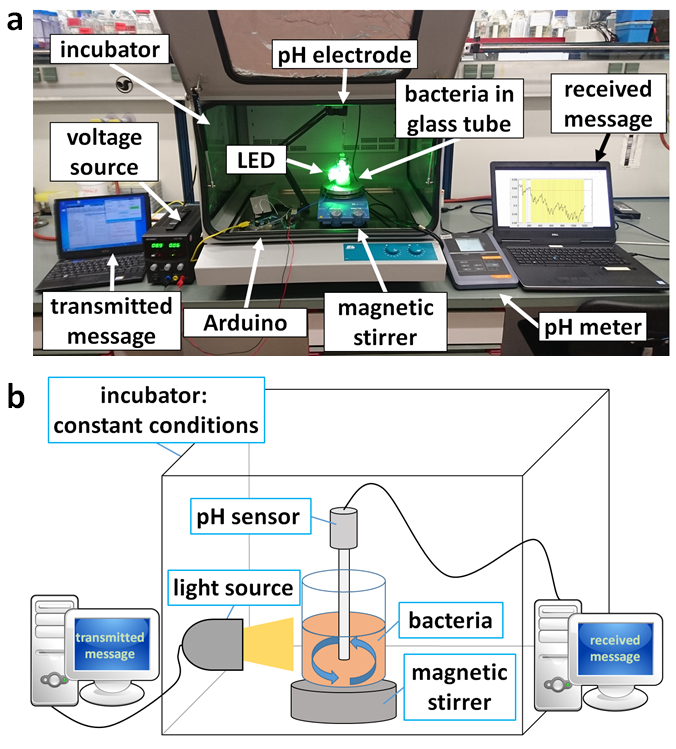}
    \caption{%
        Biological modulator model. (a) Benchtop experimental setup; (b) Schematic illustration.
    }
    \label{Fig:SysMod}
\end{figure}

\subsection{System Overview}
\begin{figure}[t]
    \includegraphics[width=0.73\columnwidth]{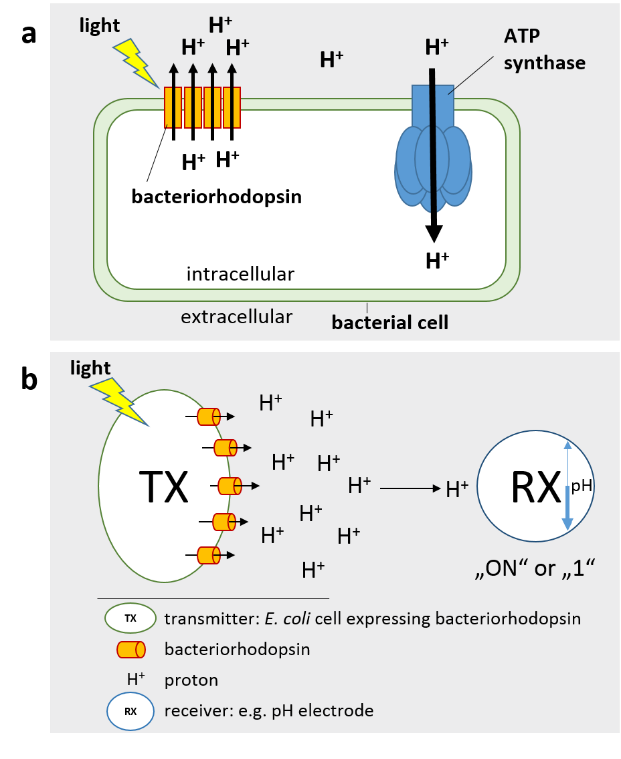}
    \caption{%
        The light-driven proton pump bacteriorhodopsin.
        (a) Biological function of bacteriorhodopsin in a native cell;
        (b) Schematic transmission model.
    }
    \label{Fig:Bacteria}
\end{figure}

The developed testbed is shown in Fig.~\ref{Fig:SysMod}a and schematically illustrated in Fig.~\ref{Fig:SysMod}b. The proposed modulator is based on \textit{E. coli} bacteria expressing bacteriorhodopsin in their cell membrane for easy-to-control optical signal conversion. A glass tube containing the bacterial suspension is installed in a light-isolated incubator in order to keep environmental conditions, such as temperature, constant. An LED is focused on the  bacterial suspension and is controlled using an Arduino microcontroller and a personal computer (PC). Thereby, the information generated by the transmitter PC is first encoded into an optical signal which is then converted to a chemical signal, i.e., a pH change, by the bacteria. In fact, upon illumination, the bacteriorhodopsin in the bacteria plasma membrane pumps protons out into the channel, see Fig.~\ref{Fig:Bacteria}. Assuming dilluted solution, the proton pumping reduces the pH according to $\text{pH}=-\log_{10}(c_{\text{H}^+})$ where $c_{\text{H}^+}$ is the concentration of protons in mol/l \cite{Farsad_Novel_2017}. In order to evaluate the efficiency of the proposed signal converter, we deploy a pH sensor in the bacterial suspension which tracks the pH variations over time. This sensor reports the pH values to a receiver PC for signal processing. The technical details of the components of the testbed and the cultivation of the bacteria are provided in Section~3 and the modulation and detection schemes used to collect and process the measurement data are presented in Section~5.

\subsection{Bacteriorhodopsin Photocycle}

The modulator in this testbed consists of engineered bacterial cells expressing GR inserted into the plasma membrane of the cell\footnote{GR is a specific bacteriorhodopsin belonging to the family of bacterial type~I rhodopsins. Throughout this paper, we use GR and bacteriorhodopsin interchangeably.}. The GR protein provides a gate through the membrane via seven transmembrane domains formed by amino acid helices. Due to a hydrophobic barrier on the cytoplasmic side, the protein is not providing any transport of molecules in the ground state. To perform proton-transfer, a chromophore group, the all-trans retinal, is needed. The retinal is integrated into the protein and acts as biochemical pumping lever. The photocycle of bacteriorhodopsin was investigated intensively over the past decades in the biology community \cite{Lanyi_Bacteriorhodopsin_2004}. In the ground state, retinal is in the all-trans configuration and a proton is bound to the residue of amino acid Asp96 inside the cell on the cytoplasmic side. By the energy of one photon, the retinal is subject to a trans$\rightarrow$cis transition at carbon atom C14 and thereby performs the lever action. As a result, one proton is transferred from the Schiff base to the residue of amino acid Asp85 on the periplasmic side of the protein. Investigations of the bacteriorhodopsin photocycle strongly suggest that the protonation of Asp85 causes the passage of a proton through a water network embedded in the amino acid residues of the protein on the extracellular side \cite{Patterson_Ultrafast_2010}. Hence, the proton can move through the plasma membrane against the electrochemical potential along the amino acid residues inside the protein. Furthermore, Asp85 reprotonates the Schiff base and the retinal regenerates to the ground state, ready for a new cycle. 

The photo-isomerization of retinal and the release of one proton to the extracellular side is the fastest known bacterial photoreaction and is performed in less than \SI{1}{\micro\second} \cite{Patterson_Ultrafast_2010}. However, the regeneration from the excited to the ground state takes 15~ms which makes it the time-limiting factor in the photocycle. By increasing the proton-gradient in the natural host, the polarization of the membrane is  used to drive, e.g., an ATPase to convert light energy to chemical energy, see Fig.~\ref{Fig:Bacteria}a, or to drive the flagellar apparatus. 

Light is one of the most important external signals used to convey information from the external world to biological systems. In fact, in addition to the ion transporting rhodopsins, there are also sensory rhodopsins functioning as light-signal transducers in nature \cite{Kaneko_Conversion_2017}. Therefore, organisms make use of light not only as energy source but also as information signal. In this paper, we exploit bacteriorhodopsin for a biological modulator.

\section{Experimental Setup}
In this section, we first describe the procedure for cultivation of the bacteria, formation of the spheroplasts that are needed for efficient proton pumping, and the performed measurement mechanism. Subsequently, we provide a brief discussion on the variability that is expected to occur in MC systems that employ biological~components. 

\subsection{Bacterial Cultivation}

In this paper, we use genetically modified \textit{E. coli} bacteria, namely the strain \textit{E. coli} $\textrm{DH5}\alpha\textrm{Mcr}$, carrying the vector DNA pKJ900 with the gene encoding GR from \textit{Gloeobacter violaceus} under control of the chemically induced \textit{ptac} promoter that was proposed in \cite{Choi_Cyanobacterial_2014}. Bacteria from a dry agar culture were pre-cultured for 6~h at \SI{37}{\celsius} and shaked at 175 revolutions per minute (rpm) in 20~ml lysogeny broth (LB) medium (i.e., 10~g/l tryptone, 10~g/l NaCl, 5~g/l yeast extract)  with \SI{25}{\micro\gram/\milli l} chloramphenicol, to select bacteria with antibiotic resistance genetically encoded in the vector DNA. Subsequently, for the main culture, 400~ml LB with chloramphenicol was inoculated to a final optical density at 600~nm of OD\textsubscript{600~nm}$=0.02$ (approximately $0.02\times (8\times 10^{8})=1.6\times 10^{7}$ cells/ml), and incubated for 1~h at \SI{37}{\celsius} at 175~rpm constant shaking, to adapt to the fresh medium conditions. Thereafter, \SI{100}{\micro M} isopropyl-$\beta$-D-thiogalactopyranosid (IPTG) for chemical induction of the transcription, and \SI{10}{\micro M} retinal were added. Since \textit{E. coli} cells do not produce retinal, it has to be supplied by the medium in which the cells grow. Afterwards, the main culture was incubated at \SI{35}{\celsius} and 75~rpm in the dark, since lower temperature supports IPTG induction and retinal incorporation seems to be more successful in rather anaerobe conditions \cite{Gopal_Strategies_2013,Hartmann_Anaerobic_1980}. 

\subsection{Spheroplast Formation}
Considering that GR is located in the plasma membrane, protons are pumped to the periplasmic space between the cytosolic and the outer membrane (OM). Therefore, most released protons are trapped and cannot easily reach the extracellular environment. To address this issue, we standardized a protocol based on sonication with the aim to remove the OM so that the protons are released directly into the surrounding medium. Among many protocols already described in the literature, using lysozyme resulted in a high and pure yield of spheroplasts but in a lower final volume and concentration of cells compared to sonication \cite{Hobb_Evaluation_2009,Liu_effect_2006}. 
Thus, OM removal by sonication is the method that better fitted the requirements of the system. The IPTG-induced cells were harvested by centrifugation (4000~xg, 5~min). Afterwards, the cells were resuspended in 320~ml 0.9\% NaCl in total and exposed to 6 times of 20~s sonication bath with ice (10\% power in Bandelin Sonorex Digital 10P) and 20~s regeneration.  After centrifugation with 8000~xg for 10~min, the cell pellet was resuspended again in 320~ml 0.9\% NaCl. The proportion of OM removal is strongly dependent on the dilution during sonication. In total, 6 cycles of sonication were performed. After the last centrifugation, the spheroplasts were resuspended in an unbuffered, osmotically balancing solution (120~mM NaCl, 10~mM MgCl$_2$, 10~mM KCl, 10~mM MgSO$_4$, \SI{100}{\micro M} CaCl$_2$) of pH=5.5 and adjusted to an OD\textsubscript{600~nm} of 15 (approximately $15\times (8\times 10^{8})=1.2\times 10^{10}$ cells/ml). The resulting solution was a mixture of spheroplasts (50-60\%, optically estimated), cells with partly removed OM, and intact cells. In a reaction tube, 6~ml of the cell suspension was incubated for 5~h in \SI{35}{\celsius} in the testbed setup, see Fig.~\ref{Fig:SysMod}, stirred at level 6 of a IKA RCT basic magnetic stirrer, and finally dark adapted to the ionic conditions before it was used for signal transmission.


\subsection{Measurement}
The bacteria, constantly incubated in a dark environment, were illuminated by an LED with optical power 1~W, which operated at wavelength 550~nm due to the maximum absorption  of GR \cite{Choi_Cyanobacterial_2014}. The LED was controlled by a custom Matlab\textsuperscript{\textregistered} (MathWorks\textsuperscript{\textregistered}, Natick, MA, United States) graphical user interface (GUI), which allowed mapping a user-defined bit sequence to an appropriate sequence of light stimuli. The transmitter PC was connected to an Arduino Mega 2560 (Rev. 3) microcontroller via serial connection. The GUI controled one of the digital output pins of the microcontroller, which in turn provided the control signal for the custom LED driver circuit PT4115 (CR Powtech, Shaghai, China). The measurement was performed when the temperature was stable at $35\pm\SI{0.2}{\celsius}$ and the pH was adapted to between 5.6-5.8, since this was the most effective operating range to generate a strong signal from the bacteria \cite{Wang_Spectroscopic_2003}. The pH signal in general was documented for at least 30 min to ensure stability. The absolute pH level was detected with a SenTix 950 (Xylem Analytics, WTW, Weilheim, Germany) microelectrode using the potentiometric pH meter inoLab\textsuperscript{\textregistered} Multi 9310 IDS (Xylem Analytics, WTW, Weilheim, Germany). Since our main objective was to characterize the optical-to-chemical signal conversion, the pH microelectrode was inserted directly into the bacterial solution. The measured real-time data were continuously streamed via serial connection to the receiver PC, where they were analyzed, displayed, and stored by a custom Matlab\textsuperscript{\textregistered} GUI. 

\subsection{Variability in Biological Systems}

The proposed testbed is based on living biological organisms. Hence, we expect conditional unique characteristics that usually are not observed in synthetic non-living systems. In particular, in the following, we highlight the factors that may cause variations in the overall system response and may help in interpreting the experimental data. These factors include, but are not limited to, the portion of spheroplasts in the cell mixture and the number of GR molecules with integrated retinal in each cell. Moreover, as bacteria age, changes in the system response could also arise from degenerating processes in the cell. These factors may lead to a baseline drift in the chemical signal over time, cf. Section~5. To minimize these effects, we followed a careful protocol for preparation of the bacteria and the  measurement procedure as discussed in the previous subsections. Nevertheless, residual variations still exist that will be studied and modeled in the next section. Determining for example the exact numbers of spheroplasts and functional GR molecules, and the development of efficient protocols to reduce or eliminate variations in these numbers, are important topics for future research.

\section{System Characterization}
In this section, we develop an analytical model to characterize the  chemical signal induced by the bacteria as a function of the applied optical signal.

\subsection{System Step Response to Illumination and Darkness}

Let $T^{\mathrm{symb}}$ denote the length of a symbol interval. In this paper, we assume a rectangular pulse for the optical signal that spans  fraction $\alpha$ of the symbol interval. In other words, for this pulse shape, the LED is turned on from the beginning of the symbol interval until time $\alpha T^{\mathrm{symb}}$ and is turned off for the remaining time $(1-\alpha) T^{\mathrm{symb}}$ of the symbol interval. Moreover, before transmission starts, the bacteria are in a dark adaptation state and  an equilibrium in pH is established. Our motivation for adopting this specific pulse shape is to partially return to the equilibrium state after illumination. This is particularly important if pulses are transmitted consecutively, e.g., corresponding to consecutive binary ones for OOK signaling.  To characterize this system, in the following, we develop an analytical model for the step response of the system to  illumination and darkness.

\subsection{Analytical Model}

Anaytical models to describe the proton release rate or proton concentration (or equivalently pH) as a function of a given induced optical intensity have been developed in \cite{Hamid_Modulator,zifarelli2008buffered}. In particular, in \cite{Hamid_Modulator}, the photocycle of the bacteriorhodopsin was modeled as a Markov chain and the corresponding proton release rate was derived. Moreover, in \cite{zifarelli2008buffered}, the expected pH change in the proximity of a proton pumping cell was derived as a function of time. In this paper, we do not aim to develop such models as the considered system is much more complex than those investigated in \cite{Hamid_Modulator,zifarelli2008buffered}. Nevertheless, we exploit a simple insight that these analytical models provide. Specifically, the model in \cite{zifarelli2008buffered} reveals an exponential change in proton concentration over time in response to the change of the optical intensity and convergence of the proton concentration to an equilibrium level after a sufficient time. We take this insight into account to develop a parametric model. Moreover, from our measurement data, we observed that the proton concentration may exhibit a certain drift that was not predicted in \cite{Hamid_Modulator,zifarelli2008buffered} but is included in our model. This effect can be attributed to a slow variation in the behavior of the bacteria over time, e.g., due to the aging of the bacteria.

Let $c_{\text{H}^+}(t)$ denote the proton concentration as a function of time. Motivated by the measurement data, we model $c_{\text{H}^+}(t)$ as the sum of the following three different components: 

\textit{i)} \textbf{Slow drift component:} As mentioned before, our measurements exhibit a slow drift over relatively long time intervals (e.g., on the order of $20$ min) compared to the considered symbol interval duration, which is on the order of $1$ min. The concentration change due to this drift is denoted by $d(t)$ and is modeled by the linear deterministic function $d(t)=m^{\mathrm{d}}t$ where $m^{\mathrm{d}}$ is the slope of the drift when measurement starts at $t=0$.

\textit{ii)} \textbf{Signal-dependent component:} A variation in the optical signal causes a change in the proton concentration within each symbol interval. Let $c_{\text{H}^+}^{\mathrm{eq},0}$ and $c_{\text{H}^+}^{\mathrm{eq},1}$ denote the equilibrium proton concentrations, i.e., $\underset{t\to\infty}{\lim}\,\,c_{\text{H}^+}(t)$, under darkness  and illumination, respectively, when the drift and noise components of $c_{\text{H}^+}(t)$ are absent. Assuming that the change of proton concentration, denoted by $x(t)$, or equivalently the change of pH level, in the bacteria suspension is proportional to the deviation from the equilibrium level, $x(t)$ can be modeled by the following ordinary differential equation (ODE):
\begin{equation}
	\frac{\mathrm{d}x(t)}{\mathrm{d}t} = -\frac{1}{\tau_i} \left(x(t) - c_{\text{H}^+}^{\mathrm{eq},i} \right),\,
\label{eq:ODE}
\end{equation}
where $\tau_0$ and $\tau_1$ are the time constants for the darkness and illumination states, respectively. Considering $x(t_0)=0$ at initial time $t_0$, the ODE in (\ref{eq:ODE}) has the following exponential solution
\begin{equation}
	x(t) = \left(c_{\text{H}^+}^{\mathrm{eq},i} - c_{\text{H}^+}(t_0)\right) \left(1-\exp\left(-t/\tau_i\right)\right). 
	\label{eq:ODE_sol}
\end{equation}
 For the pulse shape introduced in Section~4.1 and assuming that the start of the symbol interval is at time $t_0$, the proton concentration change $x(t)$ within one symbol interval is obtained as
\begin{IEEEeqnarray}{ll}
	x(t) = 
	\begin{cases}
	\big(c_{\text{H}^+}^{\mathrm{eq},1} - c_{\text{H}^+}(t_0)\big) \big(1-\exp\left(-t/\tau_1\right)\big), \,\,  t_0\leq t \leq t_0+\alpha T^{\mathrm{symb}} \\
	\big(c_{\text{H}^+}^{\mathrm{eq},0} - c_{\text{H}^+}(t_0+\alpha T^{\mathrm{symb}})\big) \big(1-\exp\left(-t/\tau_0\right)\big), \\ \hspace{3.4cm} t_0+\alpha T^{\mathrm{symb}} < t \leq t_0+T^{\mathrm{symb}}
	\end{cases} \hspace{-0.2cm}
	\label{eq:overall_sol}
\end{IEEEeqnarray}

\textit{iii)} \textbf{Random fluctuation component:} There are additional fluctuations in $c_{\text{H}^+}(t)$ which are much faster than the above two components. We model these fluctuations as noise denoted by $e(t)$. This noise may include diffusion (counting) noise, pH sensor circuitry noise, and the noise inherent to the biological machinery of the bacteria. A careful modeling of these noise sources is out of the scope of this paper. One analytical approximation that is often accurate when there are several independent noise sources is to model the overall noise as Gaussian noise. The validity of the Gaussian noise model will be investigated  in future work.

To summarize, the proton concentration is given by
\begin{equation}\label{Eq:Model}
	c_{\text{H}^+}(t) = c_{\text{H}^+}(t_0) + x(t) + d(t) + e(t).
\end{equation}
Note that conditioned on the transmitted symbols, components $x(t)$ and $d(t)$ are deterministic, whereas $e(t)$ is random. The motivation for adopting the above model comes from both the analysis in \cite{zifarelli2008buffered} and our measurement data, cf. Section~5.2. We emphasize that the above model is not directly derived based on physical laws and is in fact a parametric model whose parameters can be adjusted to fit the measurement data.

\section{Experimental Verification}
In this section, we present and analyze experimental data obtained by the proposed optical-to-chemical signal conversion interface. To this end, we first present the considered transmission scheme.

\subsection{Transmission Scheme}

Data transmission is preceded by a period of dark adaptation of \SI{30}{\minute}. Afterwards, the following modulation and detection schemes are adopted.

\subsubsection{Modulation} 

We employ OOK modulation with the pulse shape introduced in Section~4.1. Thereby, for a binary one, the LED illuminates the bacteria suspension for the first $\alpha$ fraction of the symbol interval and is turned off for the remaining fraction, whereas for a binary zero, the LED is turned off during the whole symbol interval. We note that other modulation schemes such as general concentration shift keying and pulse position modulation can be easily realized with our testbed and the analytical model proposed in Section~5 can also be straightforwardly generalized to these modulation schemes.  

\subsubsection{Detection}
Due to the random fluctuations and the drift observed in the measurement data, a simple threshold detector performs poorly for the original pH signal. To overcome these effects, we employ a smoothing filter to mitigate the random fluctuations and a differential detector to eliminate the drift. The sampling rate of the pH signal is $1$ Hz, i.e., one sample/second. The measured pH signal is smoothed by a moving average filter with a length of \SI{30}{samples} and then differentiated, where the differences were computed between signal values with a distance of \SI{20}{samples}. Then, a threshold detector is applied to recover the data from the differentiated pH signal denoted by $\Delta \text{pH}$.  The peak associated with the first binary one bit(s) can be used as synchronization signal to determine the start of transmission. The value of this peak (these peaks) can be used as a reference to determine the decision threshold. For the example shown in this section, the threshold is set as $\eta=\beta\Delta \text{pH}^{\mathrm{p}}+(1-\beta)\Delta \text{pH}^{\mathrm{d}}$ where $\Delta \text{pH}^{\mathrm{p}}$ is the peak value, $\Delta \text{pH}^{\mathrm{d}}$ is the average $\Delta \text{pH}$ before transmission starts during the dark adaptation, and $\beta\in[0,1]$ is a design parameter that determines how close the threshold is to the peak value $\Delta \text{pH}^{\mathrm{p}}$.

\begin{figure}
    \includegraphics[scale=0.55]{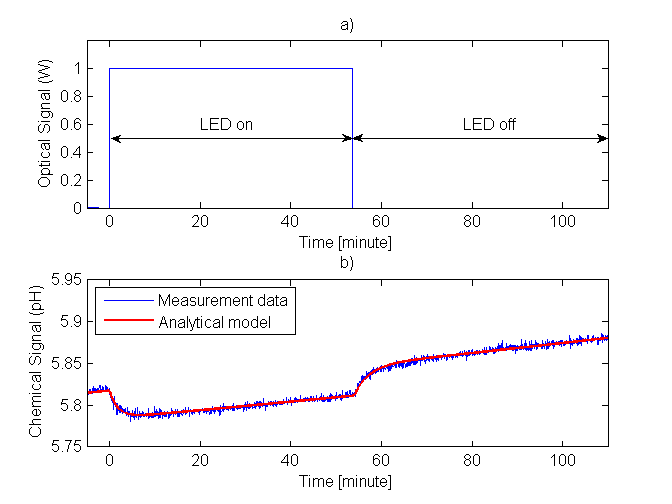}
    \caption{%
       a) Optical signal; b) Measured pH and analytical $\text{pH}^{\mathrm{mdl}}$ vs. time.
    }\label{Fig:SingleShot}
\end{figure}

\begin{figure}
    \includegraphics[scale=0.55]{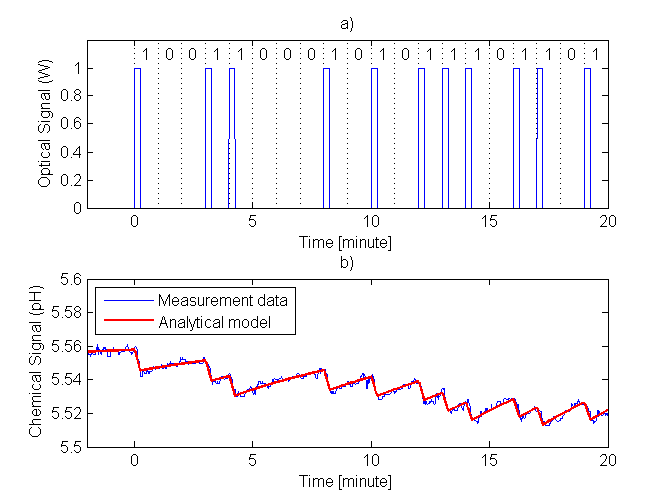}
    \caption{%
       a) Optical  signal corresponding to symbol sequence $[10011000101011101101]$; b) Measured pH and analytical $\text{pH}^{\mathrm{mdl}}$ vs. time.
    }\label{Fig:MultipleShot}
\end{figure}

\subsection{Model Verification}	

In the following, the accuracy of the model proposed in (\ref{Eq:Model}) is investigated by comparing $\text{pH}^{\mathrm{mdl}} = -\log_{10}( c_{\text{H}^+}(0) + x(t) + d(t))$ with the measurement data. We employ the least square error criterion of the MatLab Curve Fitting Toolbox\texttrademark~ to obtain model parameters $c_{\text{H}^+}^{\mathrm{eq},0}$, $c_{\text{H}^+}^{\mathrm{eq},1}$, $\tau_0$, $\tau_1$, and $m^{\mathrm{d}}$. Note that the values of these parameters may be different for different cell cultures.

In order to study the effect of the drift under both the illumination and dark states, we consider a long symbol duration of $T^{\mathrm{symb}}=110$ min with $\alpha=0.5$. In Fig.~\ref{Fig:SingleShot}a, we show the optical signal and in Fig.~\ref{Fig:SingleShot}b, we show the corresponding measured pH and the analytical $\text{pH}^{\mathrm{mdl}}$ for one cell culture  vs. time. The parameters of the proposed model are found as $c_{\text{H}^+}^{\mathrm{eq},0}=1.53\times10^{-6}$ mol/l (pH of $5.81$), $c_{\text{H}^+}^{\mathrm{eq},1}=1.65\times10^{-6}$ mol/l (pH of $5.78$), $\tau_0=3.18$ min, $\tau_1=1.84$ min, and $m^{\mathrm{d}}=-3.12\times 10^{-5}$ mol/l/s. As expected, the pH level decreases after illumination and increases during darkness; hence, the optical signal is successfully converted to a chemical signal.  From the measured pH shown in Fig.~\ref{Fig:SingleShot}b, we observe a baseline drift during both the illumination and darkness intervals. Overall, we observe that the proposed analytical model is in very good agreement with the measurement~data.

In Fig.~\ref{Fig:MultipleShot}a, we show the optical signal corresponding to the $20$-symbol sequence $[10011000101011101101]$ with $T^{\mathrm{symb}}=1$ min and $\alpha=0.25$, and in Fig.~\ref{Fig:MultipleShot}b, we show the corresponding measured pH and the analytical $\text{pH}^{\mathrm{mdl}}$  vs. time.  The parameters of the proposed model are found as $c_{\text{H}^+}^{\mathrm{eq},0}=2.82\times10^{-6}$ mol/l  (pH of $5.54$), $c_{\text{H}^+}^{\mathrm{eq},1}=5.79\times10^{-6}$ mol/l  (pH of $5.23$), $\tau_0=6.39$ min, $\tau_1=8.48$ min, and $m^{\mathrm{d}}=-6.43\times 10^{-5}$ mol/l/s. The values of the model parameters are different from those obtained for Fig.~\ref{Fig:SingleShot} since the measurements were gathered from different bacterial cultures. Again, we observe from Fig.~\ref{Fig:MultipleShot}a that the proposed analytical model explains the measurement data well even if multiple symbols are transmitted.

\subsection{Signal Conversion}

\begin{figure}
			\includegraphics[width=\columnwidth]{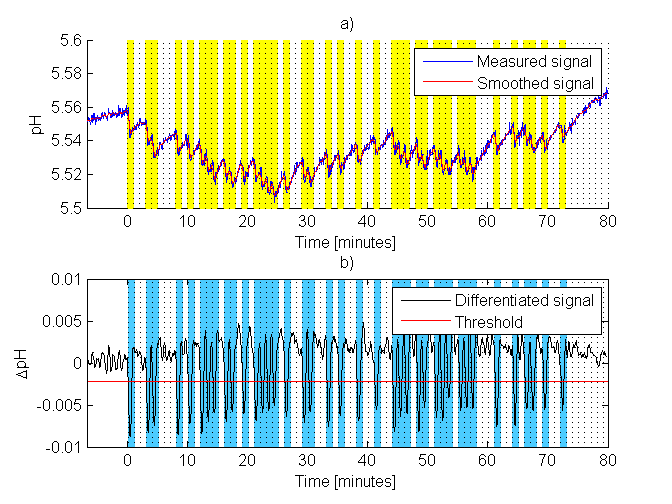}
		\caption{a) Measured pH and smoothed signal vs. time for the symbol sequence in (\ref{Eq:Sequence}); b) Differentiated signal used for detection vs. time and adopted detection threshold. The symbol intervals are represented by vertical dotted lines. For illustration, time intervals where binary symbol ``1'' is transmitted (detected) are highlighted by yellow (blue) in Fig.~\ref{Fig:Diff_detection}a (Fig.~\ref{Fig:Diff_detection}b).}%
		\label{Fig:Diff_detection}%
	\end{figure}

Finally, we show the successful recovery of the following randomly chosen $80$-symbol sequence
\begin{IEEEeqnarray}{ll}\label{Eq:Sequence}
	[10011000&10101110110101111010011001010010 \nonumber\\ 
  &0100111011011101110001001011010010000000]
\end{IEEEeqnarray}
that is converted from an optical signal to a chemical signal using the proposed experimental setup and the modulation and detection schemes introduced in Section~5.1 with $T^{\mathrm{symb}}=1$ min and $\alpha=0.25$. In Fig.~\ref{Fig:Diff_detection}a, we show the measured pH and the smoothed signal vs. time. As can be observed from this figure, the random noise in the measured pH is efficiently mitigated by smoothing. Nevertheless, different signal drifts can be observed in intervals $[0,25]$, $[26,44]$, $[45,69]$, and $[70,80]$ min which are caused by inter-symbol interference (ISI) as well as the baseline drift. Moreover, these drifts are not present in Fig.~\ref{Fig:Diff_detection}b which depicts the differentiated signal vs. time. The negative peaks in the differentiated signal, which result from illumination when transmitting a binary ``1'', are very pronounced and substantially exceed the noise level. Thereby, a simple threshold detector using a detection threshold $\eta$ with $\beta=0.25$ can successfully recover all $80$ symbols.

\section{Conclusions and Future Work}

In this paper, we introduced a biological microscale modulator based on \textit{E.~coli} bacteria that express the light-driven proton pump gloeorhodopsin and, in response to external light stimuli, can locally change their surrounding pH level by pumping protons into the channel. We provided an analytical model to characterize the induced chemical signal as a function of the applied optical signal. We further showed that the results from the proposed analytical model are in very good agreement with the measurement data for a sequence of transmitted symbols. Furthermore, using   a pH sensor as detector, employing OOK modulation, and detection based on the differential signal, a sample sequence of $80$ consecutive bits was converted to a chemical signal and successfully recovered. We note that the high data rate of at least $1$ bit/min achieved by our testbed  is a big step forward compared to existing organic testbeds (e.g., the data rate of the system in \cite{Krishnaswamy_Time_2013} is approximately $1$ bit/h). In future work, we plan to replace the pH sensor by a bacterial receiver, e.g. a pH-sensitive green fluorescent protein (GFP). Having both an optical-to-chemical transmitter and a chemical-to-optical receiver, we can set up a full MC system at microscale that can be easily controlled and read out at macroscale.



\bibliographystyle{myACM}
{\footnotesize
	\bibliography{nanocom2018}


\begin{thebibliography}{00}


\ifx \showCODEN    \undefined \def \showCODEN     #1{\unskip}     \fi
\ifx \showDOI      \undefined \def \showDOI       #1{{\tt DOI:}\penalty0{#1}\ }
  \fi
\ifx \showISBNx    \undefined \def \showISBNx     #1{\unskip}     \fi
\ifx \showISBNxiii \undefined \def \showISBNxiii  #1{\unskip}     \fi
\ifx \showISSN     \undefined \def \showISSN      #1{\unskip}     \fi
\ifx \showLCCN     \undefined \def \showLCCN      #1{\unskip}     \fi
\ifx \shownote     \undefined \def \shownote      #1{#1}          \fi
\ifx \showarticletitle \undefined \def \showarticletitle #1{#1}   \fi
\ifx \showURL      \undefined \def \showURL       #1{#1}          \fi
\providecommand\bibfield[2]{#2}
\providecommand\bibinfo[2]{#2}
\providecommand\natexlab[1]{#1}
\providecommand\showeprint[2][]{arXiv:#2}

\bibitem[\protect\citeauthoryear{Arjmandi, Jamali, Ahmadzadeh, Burkovski,
  Schober, and Kenari}{Arjmandi et~al\mbox{.}}{2016}]%
        {Hamid_Modulator}
\bibfield{author}{\bibinfo{person}{H. Arjmandi}, \bibinfo{person}{V. Jamali},
  \bibinfo{person}{A. Ahmadzadeh}, \bibinfo{person}{A. Burkovski},
  \bibinfo{person}{R. Schober}, {and} \bibinfo{person}{M.~N. Kenari}.}
  \bibinfo{year}{2016}\natexlab{}.
\newblock \showarticletitle{Ion pump based bio-synthetic modulator model for
  diffusive molecular communications}. In \bibinfo{booktitle}{{\em Proc. IEEE
  SPAWC}}. \bibinfo{pages}{1--6}.
\newblock


\bibitem[\protect\citeauthoryear{Bicen, Austin, Akyildiz, and Forest}{Bicen
  et~al\mbox{.}}{2015}]%
        {Akyildiz_testbed_2015}
\bibfield{author}{\bibinfo{person}{A.~O. Bicen}, \bibinfo{person}{C.~M.
  Austin}, \bibinfo{person}{I.~F. Akyildiz}, {and} \bibinfo{person}{C.~R.
  Forest}.} \bibinfo{year}{2015}\natexlab{}.
\newblock \showarticletitle{Efficient sampling of bacterial signal transduction
  for detection of pulse-amplitude modulated molecular signals}.
\newblock \bibinfo{journal}{{\em IEEE Trans. Biomedical Circuits and Syst.\/}}
  \bibinfo{volume}{9}, \bibinfo{number}{4} (\bibinfo{date}{Aug.}
  \bibinfo{year}{2015}), \bibinfo{pages}{505--517}.
\newblock
\showISSN{1932-4545}


\bibitem[\protect\citeauthoryear{Choi, Shi, Brown, and Jung}{Choi
  et~al\mbox{.}}{2014}]%
        {Choi_Cyanobacterial_2014}
\bibfield{author}{\bibinfo{person}{A.~R. Choi}, \bibinfo{person}{L. Shi},
  \bibinfo{person}{L.~S. Brown}, {and} \bibinfo{person}{K.-H. Jung}.}
  \bibinfo{year}{2014}\natexlab{}.
\newblock \showarticletitle{Cyanobacterial light-driven proton pump,
  Gloeobacter rhodopsin: complementarity between rhodopsin-based energy
  production and photosynthesis}.
\newblock \bibinfo{journal}{{\em PLoS One\/}} \bibinfo{volume}{9},
  \bibinfo{number}{10} (\bibinfo{year}{2014}), \bibinfo{pages}{e110643}.
\newblock


\bibitem[\protect\citeauthoryear{Farsad, Guo, and Eckford}{Farsad
  et~al\mbox{.}}{2013}]%
        {Farsad_Tabletop_2013}
\bibfield{author}{\bibinfo{person}{N. Farsad}, \bibinfo{person}{W. Guo}, {and}
  \bibinfo{person}{A.~W. Eckford}.} \bibinfo{year}{2013}\natexlab{}.
\newblock \showarticletitle{Tabletop molecular communication: Text messages
  through chemical signals}.
\newblock \bibinfo{journal}{{\em PLoS One\/}} \bibinfo{volume}{8},
  \bibinfo{number}{12} (\bibinfo{year}{2013}), \bibinfo{pages}{e82935}.
\newblock


\bibitem[\protect\citeauthoryear{Farsad, Kim, Eckford, and Chae}{Farsad
  et~al\mbox{.}}{2014}]%
        {Farsad_Channel_2014}
\bibfield{author}{\bibinfo{person}{N. Farsad}, \bibinfo{person}{N.-R. Kim},
  \bibinfo{person}{A.~W. Eckford}, {and} \bibinfo{person}{C.-B. Chae}.}
  \bibinfo{year}{2014}\natexlab{}.
\newblock \showarticletitle{Channel and noise models for nonlinear molecular
  communication systems}.
\newblock \bibinfo{journal}{{\em IEEE J. Sel. Areas Commun.\/}}
  \bibinfo{volume}{32}, \bibinfo{number}{12} (\bibinfo{year}{2014}),
  \bibinfo{pages}{2392--2401}.
\newblock


\bibitem[\protect\citeauthoryear{Farsad, Pan, and Goldsmith}{Farsad
  et~al\mbox{.}}{2017}]%
        {Farsad_Novel_2017}
\bibfield{author}{\bibinfo{person}{N. Farsad}, \bibinfo{person}{D. Pan}, {and}
  \bibinfo{person}{A. Goldsmith}.} \bibinfo{year}{2017}\natexlab{}.
\newblock \showarticletitle{A novel experimental platform for in-vessel
  multi-chemical molecular communications}. In \bibinfo{booktitle}{{\em Proc.
  IEEE Globecom}}.
\newblock


\bibitem[\protect\citeauthoryear{Farsad, Yilmaz, Eckford, Chae, and Guo}{Farsad
  et~al\mbox{.}}{2016}]%
        {Farsad_comprehensive_2016}
\bibfield{author}{\bibinfo{person}{N. Farsad}, \bibinfo{person}{H.~B. Yilmaz},
  \bibinfo{person}{A. Eckford}, \bibinfo{person}{C.-B. Chae}, {and}
  \bibinfo{person}{W. Guo}.} \bibinfo{year}{2016}\natexlab{}.
\newblock \showarticletitle{A comprehensive survey of recent advancements in
  molecular communication}.
\newblock \bibinfo{journal}{{\em IEEE Commun. Surv. Tutorials\/}}
  \bibinfo{volume}{18}, \bibinfo{number}{3} (\bibinfo{year}{2016}),
  \bibinfo{pages}{1887--1919}.
\newblock


\bibitem[\protect\citeauthoryear{Felicetti, Femminella, Reali, Gresele,
  Malvestiti, and Daigle}{Felicetti et~al\mbox{.}}{2014}]%
        {Felicetti_Modeling_2014}
\bibfield{author}{\bibinfo{person}{L. Felicetti}, \bibinfo{person}{M.
  Femminella}, \bibinfo{person}{G. Reali}, \bibinfo{person}{P. Gresele},
  \bibinfo{person}{M. Malvestiti}, {and} \bibinfo{person}{J.~N. Daigle}.}
  \bibinfo{year}{2014}\natexlab{}.
\newblock \showarticletitle{Modeling {CD40}-based molecular communications in
  blood vessels}.
\newblock \bibinfo{journal}{{\em IEEE Trans. Nanobioscience\/}}
  \bibinfo{volume}{13}, \bibinfo{number}{3} (\bibinfo{year}{2014}),
  \bibinfo{pages}{230--243}.
\newblock


\bibitem[\protect\citeauthoryear{Gopal and Kumar}{Gopal and Kumar}{2013}]%
        {Gopal_Strategies_2013}
\bibfield{author}{\bibinfo{person}{G.~J. Gopal} {and} \bibinfo{person}{A.
  Kumar}.} \bibinfo{year}{2013}\natexlab{}.
\newblock \showarticletitle{Strategies for the production of recombinant
  protein in Escherichia coli}.
\newblock \bibinfo{journal}{{\em The Protein J.\/}} \bibinfo{volume}{32},
  \bibinfo{number}{6} (\bibinfo{year}{2013}), \bibinfo{pages}{419--425}.
\newblock


\bibitem[\protect\citeauthoryear{Hartmann, Sickinger, and Oesterhelt}{Hartmann
  et~al\mbox{.}}{1980}]%
        {Hartmann_Anaerobic_1980}
\bibfield{author}{\bibinfo{person}{R. Hartmann}, \bibinfo{person}{H.-D.
  Sickinger}, {and} \bibinfo{person}{D. Oesterhelt}.}
  \bibinfo{year}{1980}\natexlab{}.
\newblock \showarticletitle{Anaerobic growth of halobacteria}.
\newblock \bibinfo{journal}{{\em Proc. National Academy of Sciences\/}}
  \bibinfo{volume}{77}, \bibinfo{number}{7} (\bibinfo{year}{1980}),
  \bibinfo{pages}{3821--3825}.
\newblock


\bibitem[\protect\citeauthoryear{Hobb, Fields, Burns, and Thompson}{Hobb
  et~al\mbox{.}}{2009}]%
        {Hobb_Evaluation_2009}
\bibfield{author}{\bibinfo{person}{R.~I. Hobb}, \bibinfo{person}{J.~A. Fields},
  \bibinfo{person}{C.~M. Burns}, {and} \bibinfo{person}{S.~A. Thompson}.}
  \bibinfo{year}{2009}\natexlab{}.
\newblock \showarticletitle{Evaluation of procedures for outer membrane
  isolation from Campylobacter jejuni}.
\newblock \bibinfo{journal}{{\em Microbiology\/}} \bibinfo{volume}{155},
  \bibinfo{number}{3} (\bibinfo{year}{2009}), \bibinfo{pages}{979--988}.
\newblock


\bibitem[\protect\citeauthoryear{Inoue, Ono, Abe-Yoshizumi, Yoshizawa, Ito,
  Kogure, and Kandori}{Inoue et~al\mbox{.}}{2013}]%
        {LightPump_2013}
\bibfield{author}{\bibinfo{person}{K. Inoue}, \bibinfo{person}{H. Ono},
  \bibinfo{person}{R. Abe-Yoshizumi}, \bibinfo{person}{S. Yoshizawa},
  \bibinfo{person}{H. Ito}, \bibinfo{person}{K. Kogure}, {and}
  \bibinfo{person}{H. Kandori}.} \bibinfo{year}{2013}\natexlab{}.
\newblock \showarticletitle{A light-driven sodium ion pump in marine bacteria}.
\newblock \bibinfo{journal}{{\em Nature Commun.\/}}  \bibinfo{volume}{4}
  (\bibinfo{year}{2013}), \bibinfo{pages}{1678}.
\newblock


\bibitem[\protect\citeauthoryear{Kaneko, Inoue, Kojima, Kandori, and
  Sudo}{Kaneko et~al\mbox{.}}{2017a}]%
        {LightPump_2017}
\bibfield{author}{\bibinfo{person}{A. Kaneko}, \bibinfo{person}{K. Inoue},
  \bibinfo{person}{K. Kojima}, \bibinfo{person}{H. Kandori}, {and}
  \bibinfo{person}{Y. Sudo}.} \bibinfo{year}{2017}\natexlab{a}.
\newblock \showarticletitle{Conversion of microbial rhodopsins: insights into
  functionally essential elements and rational protein engineering}.
\newblock \bibinfo{journal}{{\em Biophys. Rev.\/}} \bibinfo{volume}{9},
  \bibinfo{number}{6} (\bibinfo{year}{2017}), \bibinfo{pages}{861--876}.
\newblock


\bibitem[\protect\citeauthoryear{Kaneko, Inoue, Kojima, Kandori, and
  Sudo}{Kaneko et~al\mbox{.}}{2017b}]%
        {Kaneko_Conversion_2017}
\bibfield{author}{\bibinfo{person}{A. Kaneko}, \bibinfo{person}{K. Inoue},
  \bibinfo{person}{K. Kojima}, \bibinfo{person}{H. Kandori}, {and}
  \bibinfo{person}{Y. Sudo}.} \bibinfo{year}{2017}\natexlab{b}.
\newblock \showarticletitle{Conversion of microbial rhodopsins: insights into
  functionally essential elements and rational protein engineering}.
\newblock \bibinfo{journal}{{\em Biophys. Rev.\/}} \bibinfo{volume}{9},
  \bibinfo{number}{6} (\bibinfo{year}{2017}), \bibinfo{pages}{861--876}.
\newblock


\bibitem[\protect\citeauthoryear{Koo, Lee, Yilmaz, Farsad, Eckford, and
  Chae}{Koo et~al\mbox{.}}{2016}]%
        {Koo_Molecular_2016}
\bibfield{author}{\bibinfo{person}{B.-H. Koo}, \bibinfo{person}{C. Lee},
  \bibinfo{person}{H.~B. Yilmaz}, \bibinfo{person}{N. Farsad},
  \bibinfo{person}{A. Eckford}, {and} \bibinfo{person}{C.-B. Chae}.}
  \bibinfo{year}{2016}\natexlab{}.
\newblock \showarticletitle{Molecular {MIMO}: From theory to prototype}.
\newblock \bibinfo{journal}{{\em IEEE J. Sel. Areas Commun.\/}}
  \bibinfo{volume}{34}, \bibinfo{number}{3} (\bibinfo{year}{2016}),
  \bibinfo{pages}{600--614}.
\newblock


\bibitem[\protect\citeauthoryear{Krishnaswamy, Austin, Bardill, Russakow,
  Holst, Hammer, Forest, and Sivakumar}{Krishnaswamy et~al\mbox{.}}{2013}]%
        {Krishnaswamy_Time_2013}
\bibfield{author}{\bibinfo{person}{B. Krishnaswamy}, \bibinfo{person}{C.~M.
  Austin}, \bibinfo{person}{J.~P. Bardill}, \bibinfo{person}{D. Russakow},
  \bibinfo{person}{G.~L. Holst}, \bibinfo{person}{B.~K. Hammer},
  \bibinfo{person}{C.~R. Forest}, {and} \bibinfo{person}{R. Sivakumar}.}
  \bibinfo{year}{2013}\natexlab{}.
\newblock \showarticletitle{Time-elapse communication: bacterial communication
  on a microfluidic chip}.
\newblock \bibinfo{journal}{{\em IEEE Trans. Commun.\/}} \bibinfo{volume}{61},
  \bibinfo{number}{12} (\bibinfo{year}{2013}), \bibinfo{pages}{5139--5151}.
\newblock


\bibitem[\protect\citeauthoryear{Lanyi}{Lanyi}{2004}]%
        {Lanyi_Bacteriorhodopsin_2004}
\bibfield{author}{\bibinfo{person}{J.~K. Lanyi}.}
  \bibinfo{year}{2004}\natexlab{}.
\newblock \showarticletitle{Bacteriorhodopsin}.
\newblock \bibinfo{journal}{{\em Annu. Rev. Physiol.\/}}  \bibinfo{volume}{66}
  (\bibinfo{year}{2004}), \bibinfo{pages}{665--688}.
\newblock


\bibitem[\protect\citeauthoryear{Lanyi}{Lanyi}{2006}]%
        {Lanyi_Proton_2006}
\bibfield{author}{\bibinfo{person}{J.~K. Lanyi}.}
  \bibinfo{year}{2006}\natexlab{}.
\newblock \showarticletitle{Proton transfers in the bacteriorhodopsin
  photocycle}.
\newblock \bibinfo{journal}{{\em Biochimica et Biophysica Acta
  (BBA)-Bioenergetics\/}} \bibinfo{volume}{1757}, \bibinfo{number}{8}
  (\bibinfo{year}{2006}), \bibinfo{pages}{1012--1018}.
\newblock


\bibitem[\protect\citeauthoryear{Liu, Liu, and Shergill}{Liu
  et~al\mbox{.}}{2006}]%
        {Liu_effect_2006}
\bibfield{author}{\bibinfo{person}{I. Liu}, \bibinfo{person}{M. Liu}, {and}
  \bibinfo{person}{K. Shergill}.} \bibinfo{year}{2006}\natexlab{}.
\newblock \showarticletitle{The effect of spheroplast formation on the
  transformation efficiency in Escherichia coli DH5$\alpha$}.
\newblock \bibinfo{journal}{{\em J. Exp. Microbiol. Immunol\/}}
  \bibinfo{volume}{9} (\bibinfo{year}{2006}), \bibinfo{pages}{81--85}.
\newblock


\bibitem[\protect\citeauthoryear{Nakano, Eckford, and Haraguchi}{Nakano
  et~al\mbox{.}}{2013}]%
        {Nakano_Molecular_2013}
\bibfield{author}{\bibinfo{person}{T. Nakano}, \bibinfo{person}{A.~W. Eckford},
  {and} \bibinfo{person}{T. Haraguchi}.} \bibinfo{year}{2013}\natexlab{}.
\newblock \bibinfo{booktitle}{{\em Molecular communication}}.
\newblock \bibinfo{publisher}{Cambridge University Press}.
\newblock


\bibitem[\protect\citeauthoryear{Nakano, Hsu, Tang, Suda, Lin, Koujin,
  Haraguchi, and Hiraoka}{Nakano et~al\mbox{.}}{2008}]%
        {Nakano_Microplatform_2008}
\bibfield{author}{\bibinfo{person}{T. Nakano}, \bibinfo{person}{Y.-H. Hsu},
  \bibinfo{person}{W.~C. Tang}, \bibinfo{person}{T. Suda}, \bibinfo{person}{D.
  Lin}, \bibinfo{person}{T. Koujin}, \bibinfo{person}{T. Haraguchi}, {and}
  \bibinfo{person}{Y. Hiraoka}.} \bibinfo{year}{2008}\natexlab{}.
\newblock \showarticletitle{Microplatform for intercellular communication}. In
  \bibinfo{booktitle}{{\em Proc. IEEE NEMS}}. \bibinfo{pages}{476--479}.
\newblock


\bibitem[\protect\citeauthoryear{Nakano, Kobayashi, Suda, Okaie, Hiraoka, and
  Haraguchi}{Nakano et~al\mbox{.}}{2014}]%
        {Nakano_Interface_2014}
\bibfield{author}{\bibinfo{person}{T. Nakano}, \bibinfo{person}{S. Kobayashi},
  \bibinfo{person}{T. Suda}, \bibinfo{person}{Y. Okaie}, \bibinfo{person}{Y.
  Hiraoka}, {and} \bibinfo{person}{T. Haraguchi}.}
  \bibinfo{year}{2014}\natexlab{}.
\newblock \showarticletitle{Externally controllable molecular communication}.
\newblock \bibinfo{journal}{{\em IEEE J. Sel. Areas in Commun.\/}}
  \bibinfo{volume}{32}, \bibinfo{number}{12} (\bibinfo{date}{Dec.}
  \bibinfo{year}{2014}), \bibinfo{pages}{2417--2431}.
\newblock
\showISSN{0733-8716}


\bibitem[\protect\citeauthoryear{Patterson, Abela, Braun, Ganter, Pedrini,
  Pedrozzi, Reiche, and van Daalen}{Patterson et~al\mbox{.}}{2010}]%
        {Patterson_Ultrafast_2010}
\bibfield{author}{\bibinfo{person}{B. Patterson}, \bibinfo{person}{R. Abela},
  \bibinfo{person}{H. Braun}, \bibinfo{person}{R. Ganter}, \bibinfo{person}{B.
  Pedrini}, \bibinfo{person}{M. Pedrozzi}, \bibinfo{person}{S. Reiche}, {and}
  \bibinfo{person}{M. van Daalen}.} \bibinfo{year}{2010}\natexlab{}.
\newblock \showarticletitle{Ultrafast phenomena at the nanoscale: Novel science
  opportunities at the swissfel X-ray laser}.
\newblock \bibinfo{journal}{{\em Europhys. News\/}} \bibinfo{volume}{41},
  \bibinfo{number}{5} (\bibinfo{year}{2010}), \bibinfo{pages}{28--32}.
\newblock


\bibitem[\protect\citeauthoryear{Unterweger, Kirchner, Wicke, Ahmadzadeh,
  Ahmed, Jamali, Alexiou, Fischer, and Schober}{Unterweger
  et~al\mbox{.}}{2018}]%
        {TestBed_Harold}
\bibfield{author}{\bibinfo{person}{H. Unterweger}, \bibinfo{person}{J.
  Kirchner}, \bibinfo{person}{W. Wicke}, \bibinfo{person}{A. Ahmadzadeh},
  \bibinfo{person}{D. Ahmed}, \bibinfo{person}{V. Jamali}, \bibinfo{person}{C.
  Alexiou}, \bibinfo{person}{G. Fischer}, {and} \bibinfo{person}{R. Schober}.}
  \bibinfo{year}{2018}\natexlab{}.
\newblock \showarticletitle{Experimental molecular communication testbed based
  on magnetic nanoparticles in duct flow}.
\newblock \bibinfo{journal}{{\em Submitted to IEEE SPAWC\/}}
  (\bibinfo{year}{2018}).
\newblock
\showURL{%
\url{https://arxiv.org/abs/1803.06990}}


\bibitem[\protect\citeauthoryear{Wang, Sineshchekov, Spudich, and Spudich}{Wang
  et~al\mbox{.}}{2003}]%
        {Wang_Spectroscopic_2003}
\bibfield{author}{\bibinfo{person}{W.-W. Wang}, \bibinfo{person}{O.~A.
  Sineshchekov}, \bibinfo{person}{E.~N. Spudich}, {and} \bibinfo{person}{J.~L.
  Spudich}.} \bibinfo{year}{2003}\natexlab{}.
\newblock \showarticletitle{Spectroscopic and photochemical characterization of
  a deep ocean proteorhodopsin}.
\newblock \bibinfo{journal}{{\em J. Biological Chemistry\/}}
  \bibinfo{volume}{278}, \bibinfo{number}{36} (\bibinfo{year}{2003}),
  \bibinfo{pages}{33985--33991}.
\newblock


\bibitem[\protect\citeauthoryear{Zifarelli, Soliani, and Pusch}{Zifarelli
  et~al\mbox{.}}{2008}]%
        {zifarelli2008buffered}
\bibfield{author}{\bibinfo{person}{G. Zifarelli}, \bibinfo{person}{P. Soliani},
  {and} \bibinfo{person}{M. Pusch}.} \bibinfo{year}{2008}\natexlab{}.
\newblock \showarticletitle{Buffered diffusion around a spherical proton
  pumping cell: a theoretical analysis}.
\newblock \bibinfo{journal}{{\em Biophys. J.\/}} \bibinfo{volume}{94},
  \bibinfo{number}{1} (\bibinfo{year}{2008}), \bibinfo{pages}{53--62}.
\newblock


\end{thebibliography}
}

\end{document}